\documentclass[
 aip,
 jmp,
 amsmath,amssymb,
 reprint]{revtex4-1}
\usepackage{graphicx}

\begin{document}

\title{Considerations of particle definitions in the functional Schr\"{o}dinger formalism}
\author{A. M. Venditti}
\affiliation{Department of Physics, University of Toronto, 50 St. George Street, Toronto, Ontario M5S 3H4, Canada}
\email{avenditt@physics.utoronto.ca}
\author{C. C. Dyer}
\affiliation{ Department of Physical and Environmental Sciences, University of Toronto Scarborough \\
    Department of Astronomy and Astrophysics, University of Toronto, 50 St. George Street, Toronto, Ontario M5S 3H4, Canada}
\email{dyer@astro.utoronto.ca}



\begin{abstract}
It is often remarked in the literature\cite{Wald_1994} that particles in QFT on curved spacetime are akin to coordinates in general relativity and hence are physically meaningless.  This moral is given an explicit demonstration by giving the correspondence between the coordinates on phase space for a field theory and the particle number.  Usually the ambiguity in particle definitions is only as varied as the possible sets of observers on the spacetime.  However, there is a greater choice in coordinates on the phase space, especially for a field system with infinite degrees of freedom.  Hence, for one set of coordinates on the spacetime (one class of comoving observers) there are many different coordinates to choose on the phase space.  This demonstrates the true vacuousness of the concept of particles when defined as energy levels of the harmonic oscillator.  In order to give a definition of particles we must specify the apparatus that detects them.  The Unruh-Dewitt\cite{Birrell_Davies_1982} detector is one such apparatus, so we are not surprised to find that it gives a physically meaningful definition of particles. We give an explicit example on de Sitter spacetime and explain how the definition of particles as energy levels of the harmonic oscillator is meaningless even in simple cases.  This is done first by comparing the response of an Unruh-Dewitt detector to the expectation of the number operator.  Second, we demonstrate that by a choice of coordinates on phase space one can turn the Hamiltonian of a free Klein-Gordon field on FRW with flat spatial sections into a set of harmonic oscillators with \emph{time-independent} mass and frequency.  Also, we demonstrate a new method for determining the wave functional of known states such as the conformal vacuum.  
\end{abstract}
\maketitle
\section{Introduction}
It is well known that when doing QFT in a curved background in the Heisenberg picture there are several different complete sets of solutions one can expand the field operators in.  Thus we could have
\begin{eqnarray}
\phi(x) &=& \int{d^3k\left(u_{\vec{k}}a_{\vec{k}} + u^*_{\vec{k}}a^{\dagger}_{\vec{k}}\right)}\\
        &=& \int{d^3k\left(\bar{u}_{\vec{k}}\bar{a}_{\vec{k}} + \bar{u}^*_{\vec{k}}\bar{a}^{\dagger}_{\vec{k}}\right)}
\end{eqnarray}
where $u_{\vec{k}}$ and $\bar{u}_{\vec{k}}$ are two different solutions to the Klein-Gordon equation and $a_{\vec{k}}$, $\bar{a}_{\vec{k}}$ are two different annihilation operators.  In virtually all cases, a complete set of modes ($u_{\vec{k}}$, $u^{\dagger}_{\vec{k}}$) is chosen so that they are `natural' solutions in some coordinate system $x^c$ chosen on the background spacetime.  Here `natural' means any coordinate dependent criteria used to single out solutions.  For instance, the solutions might be the separable solutions to the Klein-Gordon equation in one or more coordinates chosen on the background spacetime (i.e. $u_{\vec{k}} = A(x^0)B(x^1)C(x^2)D(x^3)$).\\

If there exists a time-like (conformal) Killing vector field on the spacetime then it is natural to choose coordinates so that the (C.)K.V. is given by $\partial_{x^{0}}$.  One can then choose the modes that are `positive frequency' with respect to $\partial_{x^{0}}$. 
\begin{equation}
\mathcal{L}_{x^0}u_{\vec{k}} = -i\omega u_{\vec{k}}, \;\;\;\; \omega > 0
\end{equation}  
where $\mathcal{L}_{x^0}$ is the Lie derivative with respect to the time $x^{0}$.\\ 

Hence, in the usual representation of the ambiguity of particle definitions, the choice of solutions is `induced' by coordinate transformations on the spacetime.  Coordinate systems are usually tied to observers on the spacetime, so for each observer we will have a natural definition of particle states.  In this sense the definitions of particle states are only as varied as the sets of observers in the spacetime.\\

This need not be the case as is apparent when one looks at things in the Schr\"{o}dinger formalism.  There is far more freedom in the choice of coordinates on phase space than on the spacetime.  This is especially apparent when considering a field theory which has infinite degrees of freedom as opposed to the spacetime which only has four dimensions.  \\

In section \ref{sec:Minkspace}, it will be shown that the definition of particles as energy levels of the harmonic oscillator is equivalent to picking coordinates on phase space.  It will also be shown that the calculation of the Wightman function and hence the response of an Unruh-Dewitt detector will not depend on the way coordinates are chosen on phase space.  By the principle of general covariance this demonstrates that the true (physically accurate) definition of particles is given by the response of the Unruh-Dewitt detector and that the expectation of the number operator for energy levels of the harmonic oscillator is in general physically meaningless.  This is an explicit realization of a statement made in the literature\cite{Wald_1994} (see pg. 15 and pg. 60).\\

In section \ref{sec:desitterspace} an explicit example is given to demonstrate the points made in section \ref{sec:Minkspace}.  The particle expectation number of a conformal scalar field in the conformal vacuum seems to indicate that there is not a thermal distribution of particles and rather that the distribution becomes thermal as $t \rightarrow \infty$.  Despite this indication, it is shown that a comoving observer will in fact see a thermal distribution of particles for all time.\\

Also, in section \ref{sec:desitterspace} we employ a new method to determine the wave functional corresponding to the conformal vacuum.  This method is more reliable than the method found elsewhere \cite{Hill_1985} as it does not rely on how one renormalizes the stress energy tensor operator.\\

In section \ref{sec:four} it is shown that a canonical transformation can be done on phase space so that the Hamiltonian of a free Klein-Gordon field can be put in the form of a set of harmonic oscillators with \emph{constant} mass and frequency.  From this we obtain an obvious definition of particles based on energy levels of the harmonic oscillator that is different from those found elsewhere \cite{Krauss_2007} \; \cite{Greenwood_2009} \; \cite{Greenwood_2010}.  We will explain how the results of this section are at odds with some results in the literature\cite{Krauss_2007} \; \cite{Greenwood_2009} \; \cite{Greenwood_2010}.
\section{Particles in Minkowski Space} \label{sec:Minkspace}

The massless, real scalar field is usually quantized by expressing it as the following expansion
\begin{equation} \label{eq:phiexpansion1}
\phi(t, \vec{x}) = \int{\frac{d^3k}{(2\pi)^{3/2}}b_{\vec{k}}(t)\exp(i\vec{k}\cdot\vec{x})}
\end{equation}
where the $b_{\vec{k}}(t)$ are complex functions of time that must satisfy $b_{\vec{k}}^{*} = b_{-\vec{k}}$ in order for $\phi$ to be real.  It is then substituted into the action 
\begin{equation} \label{eq:action}
S = \int{d^4x \sqrt{-g}\phi_{|a}\phi_{|b}g^{ab}}
\end{equation}
with $g_{ab}$ taken to be the flat spacetime metric.  Above the notation $_{|a} \equiv \partial_{a}$ was used.  One then treats the $b_{\vec{k}}$'s as the generalized coordinates that must be quantized via the canonical commutation relations.  Quantizing the $b_{\vec{k}}$'s will lead to the ``usual'' results in Minkowski spacetime; the Poincare invariant vacuum will have a zero expectation value for the number operator of the $b_{\vec{k}}$ particles.  In flat spacetime the zero expectation of the number operator corresponds to the zero response rate of Unruh-Dewitt detectors traveling along inertial trajectories.\\

Substituting (\ref{eq:phiexpansion1}) into (\ref{eq:action}) in Minkowski spacetime in the usual Cartesian coordinates we obtain
\begin{equation}
S = \frac{1}{2}\int{dt\int{d^3k\left(|\dot{b}_{\vec{k}}(t)|^2 - |\vec{k}|^2|b_{\vec{k}}(t)|^2\right)}}
\end{equation}
where the dot over the $b_{\vec{k}}(t)$ denotes differentiation with respect to $t$.  This is the action for an infinite set of complex $b_{\vec{k}}(t)$'s.  For our purposes we can simply deal with the real part of each $b_{k}(t)$ one mode at a time.  The action for this single real-variable is
\begin{equation}
S = \frac{1}{2}\int{dt\left((\dot{x}(t))^2 - \omega^2 (x(t))^2\right)}
\end{equation}
where $\omega = |\vec{k}|$ and $x(t)$ is understood to be the real part of a single mode $b_{k}(t)$.  The Hamiltonian for the system is
\begin{equation}
H = \frac{1}{2}\left( p^2 + \omega^2x^2\right)
\end{equation}
where $p = \dot{x}$.  At this point we remark that this is a harmonic oscillator with a fixed frequency and mass.  Coordinates will be chosen on phase space so that the new Hamiltonian takes the form of a harmonic oscillator with \emph{time-dependent} mass and frequency.  The new Hamiltonian will have a natural interpretation for particle states (certainly as natural as in de Sitter spacetime as will be shown below).\\

We will perform a type 2 canonical transformation taking the coordinates $(x, p)$ to $(Q, R)$.  The generating function for this transformation is
\begin{equation}  
G_{2}(x, R) = f(t)xR + f(t)\frac{x^2}{2} + h(t)\frac{R^2}{2}
\end{equation}
where the generating function is assumed to be a function of the old coordinate $x$ and the new momenta $R$ and is explicitly a function of time through $f(t)$ and $h(t)$.  The relations giving the coordinate transformations are
\begin{eqnarray}\label{eq:canonical1}
p &=& \frac{\partial G_{2}}{\partial x} = f(t)(R + x) \\
Q &=& \frac{\partial G_{2}}{\partial R} = f(t)x + h(t)R \label{eq:canonical2}
\end{eqnarray}
The new Hamiltonian is given by
\begin{eqnarray}
H' &=& H + \frac{\partial G_{2}}{\partial t} \\
   &=& \left[ \frac{1}{2}(f-h)^2 + \frac{\omega^2}{2}\frac{h^2}{f^2} - \frac{\dot{f}}{f}h + \frac{\dot{f}h^2}{2f^2} + \frac{\dot{h}}{2}\right]R^2 \\
\nonumber   &+& \left[ \frac{1}{2} + \frac{\omega^2}{2f^2} + \frac{\dot{f}}{2f^2}\right]Q^2 \\
\nonumber   &+& \left[ (f-h) - \frac{\omega^2}{f^2}h + \frac{\dot{f}}{f} - \frac{\dot{f}}{f^2}h\right]QR
\end{eqnarray}
where we have expressed everything in the new $Q$, $R$ coordinates by applying the relations (\ref{eq:canonical1}) and (\ref{eq:canonical2}).\\

$H'$ should have the form of a harmonic oscillator.  Given that $Q$ is the new `position' and $R$ it's conjugate momentum we require that the coefficient of the $QR$ term go to zero.  The following relation is obtained between $f$ and $h$.
\begin{equation}
h = \frac{f^3 + \dot{f}f}{f^2 + \omega^2 + \dot{f}}\label{eq:hfrelation}
\end{equation}
$H'$ is now given by
\begin{equation}
H' = \frac{1}{2}\left(\frac{R^2}{M(t)} + M(t)\Omega^2(t)Q^2\right)
\end{equation}
where $M(t)$ and $\Omega(t)$ are defined appropriately.  $H'$ is the Hamiltonian of a harmonic oscillator with time-dependent frequency and mass.\\

To quantize this system we note that $[\hat{x},\hat{p}] = [\hat{Q},\hat{R}] = i$.\\

We can come up with a natural particle interpretation by specifying the annihilation and creation operators for this system.  From now on we deal with quantum operators so we drop the hats.  $\gamma(t)$ is the annihilation operator and $\gamma^{\dagger}(t)$ the creation operator with $[\gamma(t), \gamma^{\dagger}(t)] = 1$ holding.  We have that
\begin{equation}\label{eq:fakeparticles}
\gamma(t) = \frac{R}{\sqrt{2M(t)\Omega(t)}} - i\sqrt{\frac{M(t)\Omega(t)}{2}}Q
\end{equation}
The Hamiltonian can be written as
\begin{equation}
H' = \Omega(t)\left(\gamma^{\dagger}(t)\gamma(t) + \frac{1}{2}\right)
\end{equation}
We can interpret $\gamma^{\dagger}(t)$ to create a particle of energy $\Omega(t)$.\\

For purposes of comparing the above particle picture with the response of an Unruh-Dewitt detector, we wish to identify the wave function for the usual Minkowski vacuum state in the new $Q, R$ coordinates.  The old Hamiltonian can be expressed as
\begin{equation}
H = \frac{1}{2}\omega\left(a^{\dagger}a + \frac{1}{2}\right)
\end{equation}
with $a = \frac{p}{\sqrt{2\omega}} - i\sqrt{\frac{\omega}{2}}x$.  The usual Minkowski vacuum state is the ground state of each of the harmonic oscillators of the KG field.  For a single mode $a|\Psi(t)> \;\; = 0$.  We express the annihilation operator $a$ in terms of the $(Q,R)$ coordinates by using the relations (\ref{eq:canonical1}) and (\ref{eq:canonical2}).  Since we wish to find the wave function we go to the `position' space representation of the operators $Q$ and $R$.
\begin{equation}
\hat{Q} \rightarrow Q \;\;\;\;\;\; \hat{R} \rightarrow -i\frac{\partial}{\partial Q}
\end{equation}
The relation $a|\Psi(t)> \;\; = 0$ becomes in the $(Q,R)$ coordinates, in terms of the wave function
\begin{equation}
\label{eq:minkrelation} \frac{\partial \Psi(t)}{\partial Q} + A(t)Q\Psi(t) = 0
\end{equation}
where
\begin{equation}
\label{eq:Aequation} A(t) \equiv \frac{\left(\frac{1}{\sqrt{2\omega}} - i\sqrt{\frac{\omega}{2}}\frac{1}{f}\right)}{\left(\sqrt{\frac{\omega}{2}}\frac{h}{f} - i\frac{f-h}{\sqrt{2\omega}}\right)}
\end{equation}
Integrating (\ref{eq:minkrelation}) we obtain the following wave function
\begin{equation}\label{eq:minkansatz}
\Psi(t) = \exp\left(\frac{-A(t)Q^2}{2} + B(t)\right)
\end{equation}
where the function $B(t)$ is an arbitrary function of integration (it is independent of $Q$).  It can be determined up to a constant by substitution of (\ref{eq:minkansatz}) into the Schr\"{o}dinger equation
\begin{equation}
H'\Psi(t,Q) = i\frac{\partial \Psi(t, Q)}{\partial t}
\end{equation}
It can be shown that the ansatz (\ref{eq:minkansatz}) is a solution to the above Schr\"{o}dinger equation.  It thus represents the usual vacuum state in the new coordinates ($Q$, $R$).\\

We can now evaluate the expectation number of the particles, as defined by (\ref{eq:fakeparticles}), in a given mode using the relation
\begin{equation}
<N_{\vec{k}}> = \frac{\int{da_{\vec{k}} \Psi^{*}(t) \gamma^{\dagger}_{\vec{k}}(t')\gamma_{\vec{k}}(t') \Psi(t)}}{\int{da_{\vec{k}} \Psi^{*}(t) \Psi(t)}}
\end{equation}
where we have taken the state $\Psi(t)$ and the particles $\gamma(t')$ at different times.  This is done to account for the fact that the energy of particles as defined in (\ref{eq:fakeparticles}) changes with time.  One may want to take the particle energies to be constant (evaluated at a single time) while the particle number is evaluated at the current time $t$.  We find the following expression for particle number
\begin{equation}
<N_{\vec{k}}> = \frac{\Re(A(t))}{4M(t')\Omega(t')} - \frac{(\Im(A(t)))^2}{4M(t')\Omega(t')\Re{A(t)}} + \frac{M(t')\Omega(t')}{4} - \frac{1}{2}
\end{equation}
where $\Re$ and $\Im$ are the real and imaginary part respectively.  It can be checked that this expectation is in general non-zero by picking the function $f(t)$ arbitrarily (say $f(t) = e^t$).\\

We have thus demonstrated that the definition of particles as energy levels of the harmonic oscillator is akin to picking different coordinates on the phase space.  Therefore, particles defined this way are in general unphysical.  They do not correspond to what any observer would detect (not even an inertial observer with proper time t).  Any single class of observers would have several different definitions of particles based on the energy levels of the Harmonic oscillator, were one to use this as a definition of particles.

\subsection{Response of Unruh-Dewitt detector}
There is a physically meaningful concept of a ``particle''.  It is based on the response rate of the Unruh-Dewitt detector.  When the detector makes a transition to an excited state, a particle is said to be detected.  The response of the detector is given by the integral of the two point function of the scalar field.  The scalar field operator $\phi(x)$ is unaffected by any canonical coordinate transformation on phase space because canonical coordinate transformations leave the Poisson bracket invariant.  The Dirac procedure for quantization given by 
\begin{equation}
[\hat{f},\hat{g}] = \hat{\{f,g\}}
\end{equation}
guarantees that the $\phi(x)$ field is invariant under change of coordinates on phase space and hence the response of an Unruh-Dewitt detector is invariant under the transformations given above.\\

By the principle of general covariance we know that the response of a detector is a physically meaningful effect as opposed to the expectation of the number operator derived above.  It is expected that the response of the detector following a given trajectory should have a coordinate independent meaning as it corresponds to the transition rate of a physical two-level system such as an electron orbiting a hydrogen atom.\\

The response of a detector for an inertial observer in the Minkowski vacuum ($\Psi$) is known to be null \cite{Birrell_Davies_1982}. Contrast this with the generally non-zero particle number as specified by the expectation of the number operator above which is also in the Minkowski vacuum ($\Psi$).  The point here is that when the response of a detector gives us a different particle number than the expectation of the number operator we should trust the detector.

\section{Particles in de Sitter Spacetime}\label{sec:desitterspace}

In this section we will quantize a conformally-coupled, massless, real scalar field in a de Sitter spacetime background and develop a mathematical notion of particles equivalent to that found in the literature \cite{Krauss_2007}\; \cite{Greenwood_2009}\; \cite{Greenwood_2010}.  It will be shown that the expectation of the number operator appears to indicate a non-thermal distribution of particles that becomes thermal as $t \rightarrow 0$ where t is the conformal time.  It is false to conclude that this would correspond to what a comoving observers sees or that the state is not a thermal one.  It will be shown that the Unruh-Dewitt detector for a comoving, inertial observer will click as if in a bath of thermal radiation and hence does not agree with the expectation of the particle number operator.  \\

The action for a massless, conformally coupled scalar Klein-Gordon field is
\begin{equation}
S = \frac{1}{2}\int{d^4x \sqrt{-g}\left(\phi_{|a}\phi_{|b}g^{ab} - \frac{1}{6}R\phi^2 \right)}
\end{equation}
where $R$ is the Ricci scalar.  The metric for the FLRW class of spacetimes that have flat spatial secitons (of which de Sitter is a special case) in conformal time and comoving coordinates is 
\begin{equation} \label{eq:flatfrw}
ds^2 = a^2(t)\left(dt^2 - dx^2 - dy^2 - dz^2 \right)
\end{equation}
Now we will expand the scalar field using the same expansion as above in section \ref{sec:Minkspace}

\begin{equation} \label{eq:desitterbasis}
\phi(t, \vec{x}) = \int{\frac{d^3k}{(2\pi)^{3/2}}\alpha_{\vec{k}}(t)\exp(i\vec{k}\cdot\vec{x})}
\end{equation}
Substituting this into the action above with the background given by the metric (\ref{eq:flatfrw}) we obtain

\begin{equation}
S = \frac{1}{2}\int{dt d^3k \left[ a^2(t) \dot{\alpha}_{-\vec{k}}\dot{\alpha}_{\vec{k}} - \omega_{k}^2(t)\alpha_{\vec{k}}\alpha_{-\vec{k}} \right] }
\end{equation}
with $\omega_{k}^2(t) = a^2(t)k^2 + \frac{1}{6}a^4(t)R(t)$.  Here the Ricci scalar is denoted to be $R(t)$ because it will only be a function of time for the metric (\ref{eq:flatfrw}).  Now we find the momenta conjugate to the coordinate $a_{\vec{k}}$ which is given by the following formula

\begin{equation}
\Pi_{\vec{k}} = \frac{\delta S}{\delta \dot{\alpha}_{\vec{k}}} = a^2(t)\dot{\alpha}_{-\vec{k}}
\end{equation}
The Hamiltonian is given via the Legendre transformation, with the result

\begin{eqnarray}
H &=& \int{d^3 k \Pi_{\vec{k}}\dot{\alpha}_{\vec{k}}} - L \\
\nonumber  &=& \frac{1}{2}\int{d^3k \left[ \frac{1}{a^2(t)} \Pi_{\vec{k}}\Pi_{-\vec{k}} + \omega_{k}^2(t)\alpha_{\vec{k}}\alpha_{-\vec{k}} \right] } \label{eq:frwhamiltonian}
\end{eqnarray}
The canonical commutation relations are given by

\begin{equation}
[\hat{\alpha}_{\vec{k}}, \hat{\Pi}_{\vec{q}}] = i \delta(\vec{k}-\vec{q})
\end{equation}
with the following adjoint relations holding

\begin{equation}
\hat{\Pi}^{\dagger}_{\vec{q}} = \hat{\Pi}_{-\vec{q}} \;\;\;\;\;\;\; \hat{\alpha}^{\dagger}_{\vec{k}} = \hat{\alpha}_{-\vec{k}}
\end{equation}
The usual implementation satisfies the above relations.  It is given by
\begin{equation}
\hat{\alpha}_{\vec{k}} \rightarrow \alpha_{\vec{k}} \;\;\;\;\;\;\;\; \hat{\Pi}_{\vec{k}} \rightarrow -i\frac{\partial}{\partial \alpha_{\vec{k}}}
\end{equation}

Now we will solve the Schr\"{o}dinger equation for the wave functional of the scalar field.  For this case the Schr\"{o}dinger equation becomes

\begin{equation}
\frac{1}{2}\epsilon \sum_{\vec{k}}{\left[ \frac{-1}{a^2(t) \epsilon^2} \frac{\partial^2 \Psi}{\partial \alpha_{\vec{k}} \alpha_{-\vec{k}}}+ \omega_{k}^2(t)\alpha_{\vec{k}}\alpha_{-\vec{k}}\Psi\right]} = i \frac{\partial \Psi}{\partial t}
\end{equation}  
where we are working in a box or in the discrete momentum limit.  $\epsilon$ is a dimensionful parameter that represents the spacing between consecutive values of the momenta and is taken to zero in the continuum limit.  To solve the equation we use the following ansatz
\begin{equation} \label{eq:ansatz2}
\Psi = \exp \left(-\epsilon \sum_{\vec{q}}{ \left[ a_{\vec{q}}a_{-\vec{q}}(f_{q}(t) + i g_{q}) + \frac{\gamma_{q}(t) + i\delta_{q}(t)}{\epsilon} \right] } \right)
\end{equation}
In this case the functions $f_{q}(t)$, $g_{q}(t)$, $\gamma_{q}(t)$ and $\delta_{q}(t)$ are real and only depend on the magnitude of $\vec{q}$.  Substituting this into the Schr\"{o}dinger equation for this scenario we obtain the following equations

\begin{eqnarray}
\dot{\gamma}_{k} &=& \frac{-g_{k}}{a^2(t)} \\
\dot{\delta}_{k} &=& \frac{f_{k}}{a^2(t)} \\
\label{eq:fequation}
\dot{f}_{k} &=& \frac{4 g_{k} f_{k} }{ a^2(t) } \\ 
\label{eq:gequation}
\dot{g}_{k} &=& \frac{-2 f_{k}^{2} + 2g_{k}^{2} }{a^2(t)} + \frac{\omega^2_{k}(t)}{2} 
\end{eqnarray}
The above equations for $f_{k}$ and $g_{k}$ can be solved for the case of de Sitter spacetime with flat spatial sections given by $a(t) = C/t$ and $R = 12/C^2$.  To solve them use equation (\ref{eq:fequation}) to solve for $g_{k}$ and $\dot{g}_{k}$ in terms of $f_{k}$ and its derivatives and then substitute the result into equation (\ref{eq:gequation}).  This gives the following equation for $f_{q}$
\begin{equation}
\frac{\ddot{f}_{k}}{\dot{f}_{k}} - \frac{3}{2}\frac{\dot{f}_{k}}{f_{k}} - \frac{2}{t} = - \frac{8t^4f_{k}^3}{C^4\dot{f}_{k}} + \frac{4k^2f_{k}}{\dot{f}_{k}} + \frac{8f_{k}}{t^2\dot{f}_{k}}
\end{equation}
Now we make the substitution $f_{k} = (t\chi_{k})^{-2}$ and obtain the following equation for $\chi_{k}$.
\begin{equation}
\ddot{\chi_{k}} + \left( 2k^2 - \frac{2}{t^2} \right)\chi_{k} = \frac{4}{C^4\chi_{k}^3}
\end{equation}
This equation has the general solution \cite{pinney_1950}
\begin{equation}
\chi_{k} = \left( u^2(t) + \frac{4}{C^4}\frac{v^2(t)}{W^2} \right)^{1/2}
\end{equation}
where $u(t)$ and $v(t)$ are two independent solutions to the equation $\ddot{\chi_{k}} + \left( 2k^2 - 2/t^2 \right)\chi_{k} = 0$ and $W$ is their Wronskian.  The initial conditions on these solutions are $u(t_{0}) = \chi_{0}$, $\dot{u}(t_{0}) = \dot{\chi}_{0}$, $v(t_{0}) = 0$ and $ \dot{v}(t_{0}) \neq 0$.  The equation $\ddot{\chi_{k}} + \left( 2k^2 - 2/t^2 \right)\chi_{k} = 0$ can be solved in general in terms of sines, cosines and $t$.  Therefore we can write down the closed form expression for any state represented by the ansatz for $\Psi$ above.
\\

An exact solution in which we will be interested is the following
\begin{equation} \label{eq:confvacuum}
f_{\vec{q}}(t) = \frac{qC^2}{2t^2} \;\;\;\;\;\;\;\;\;\; g_{\vec{q}}(t) = \frac{-C^2}{2t^3}
\end{equation}
It will be shown below that the wave functional represented by the ansatz (\ref{eq:ansatz2}) with the above choice for the functions $f_{q}$ and $g_{q}$ is the conformal vacuum\cite{Birrell_Davies_1982} of the de Sitter spacetime.\\

\subsection{Particle Interpretation}

Now we develop a particle interpretation of the theory.  The Hamiltonian can be expressed as in equation (\ref{eq:frwhamiltonian})
\begin{equation}
H = \frac{1}{2}\int{d^3k \left[ \frac{1}{a^2(t)}\Pi_{\vec{k}}\Pi_{-\vec{k}} + \omega^2_{k}(t)\alpha_{\vec{k}}\alpha_{-\vec{k}} \right] }
\end{equation}
which is of the form of a harmonic oscillator with time-dependent mass and frequency.  Therefore it is possible to develop a particle interpretation.  Define the annihilation and creation operators respectively, by:

\begin{eqnarray}
b_{\vec{k}} &=& \sqrt{\frac{\epsilon}{2}}\left( \sqrt{a(t)\omega_{k}(t)}\hat{\alpha}_{\vec{k}} + i \frac{\hat{\Pi}_{-\vec{k}}}{\sqrt{a(t)\omega_{k}(t)}} \right) \\
b_{\vec{k}}^{\dagger} &=& \sqrt{\frac{\epsilon}{2}}\left( \sqrt{a(t)\omega_{k}(t)}\hat{\alpha}_{-\vec{k}} - i \frac{\hat{\Pi}_{\vec{k}}}{\sqrt{a(t)\omega_{k}(t)}} \right) 
\end{eqnarray}

The above annihilation/creation operators satisfy the usual commutation relations $[b_{\vec{q}}, b_{\vec{k}}^{\dagger}] = \delta_{\vec{q},\vec{k}}$.  The Hamiltonian can be written in terms of these operators as

\begin{equation}
H = \int{d^3k \frac{\omega_{k}(t)}{a(t)}\left( b^{\dagger}_{\vec{k}}b_{\vec{k}} + 1 \right)}
\end{equation}
Therefore $b^{\dagger}_{\vec{k}}$ creates particles of energy $\omega_{k}(t)/a(t)$ at time $t$.  As before we can calculate the expectation of the number operator for particles of type $b_{\vec{k}}$.  It is given by the formula

\begin{eqnarray} \label{eq:particlenumber}
<N_{\vec{k}}> &=& \frac{\int{d\alpha_{\vec{k}} \Psi^{*}b^{\dagger}_{\vec{k}}b_{\vec{k}}\Psi}}{\int{d\alpha_{\vec{k}}\Psi^*\Psi}} \\
\nonumber &=&  \frac{g_{k}^2}{a(t)\omega_{k}(t)f_{k}} + \frac{f_{k}}{a(t)\omega_{k}(t)} + \frac{\omega_{k}a(t)}{4f_{k}}  - 1 
\end{eqnarray}
The above expression gives the number of particles as defined by the creation/annihilation operators above.  This definition of particles in terms of the creation and annihilation operators with time dependent mass and frequency above is equivalent to the defintion of particles found in the literature \cite{Krauss_2007}\; \cite{Greenwood_2009}\; \cite{Greenwood_2010} as will be shown in the appendix.  The expectation number is more easily calculated by using the annihilation/creation operators as opposed to dealing with the wave functions of the multi-particle states directly.\\

Figure 1 shows a plot of number vs. frequency in the conformal vacuum state given by (\ref{eq:ansatz2}) and (\ref{eq:confvacuum}).  The expectation rises to infinity at $\nu = 0$ and rises for all $\nu$ as $t \rightarrow 0$ where t is the conformal time (equivalently as the comoving time goes to $\infty$).  The behavior of this plot in time is akin to the equivalent plots found in the literature \cite{Krauss_2007} \; \cite{Greenwood_2010}.  There the authors conclude that this indicates a state that is becoming thermal as time goes to infinity.  Below the response of an Unruh-Dewitt detector in the conformal vacuum is calculated and is found to give a response as if it was in a bath of thermal radiation.  So we see that the expectation of the number operator gives a false particle number count and it is this fact that is at odds with the results elsewhere  \cite{Krauss_2007} \; \cite{Greenwood_2010} because of the definition of particles used in these papers.\\

\begin{figure}[h!]
\centering
\caption{Number of particles vs frequency for various conformal times (t)}
\includegraphics[width = 90mm, height = 70mm]{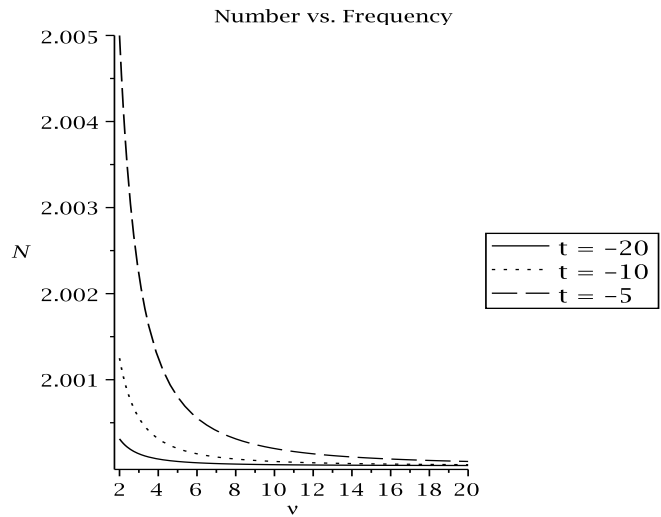}
\end{figure}

\subsection{Response of Unruh-Dewitt detector}

It was shown in section (\ref{sec:Minkspace}) that the expectation of the particle number operator is a poor indicator of what some observer actually detects with their particle detector.  Therefore we must explicitly calculate the response rate of a particle detector traveling along the observer's trajectory in order to know what the observer actually sees.  In this subsection we will have to identify the conformal vacuum wave functional in order to compare the expectation of particle numbers (as calculated above) with the response of a particle detector.  To find the wave functional we employ a new procedure where one compares the expressions for the two point function obtained in the functional Schr\"{o}dinger formalism with that in the normal Heisenberg picture.  \\

The transition rate of an Unruh-Dewitt detector at time $\tau$ is given by \cite{Schlicht_2003}

\begin{equation} \label{eq:responserate}
\dot{F}_{\tau}(\nu) = 2\lim_{\epsilon \rightarrow 0}{\int_{-\infty}^{0}{ds \,\,\, Re(\,\, \exp(i\nu s)G^{+}_{\epsilon}(x(\tau), x(\tau - s)))}}
\end{equation}
where $G^{+}_{\epsilon}(x(\tau), x(\tau - s))$ is the two point function evaluated at the two specified points on the world line and regulated by the small parameter $\epsilon$.  The formula for the two point function in the interaction picture is given by
\begin{equation}
G^{+}(x, x') = <\!\!\phi(x)\phi(x')\!\!>
\end{equation}
where $\phi(x)$ is the usual time-dependent operator that can be expanded in terms of creation and annihilation operators.\\

In the functional Schr\"{o}dinger formalism the two point function can be derived as follows.  In the interaction picture the scalar field operator at all times is given in terms of the operator at a single time as
\begin{equation}
\hat{\phi}(t,\vec{x}) = U^{\dagger}(t,t_{0})\hat{\phi}(t_{0}, \vec{x})U(t,t_{0})
\end{equation}
where $U(t,t_{0}) = \mathcal{T}\exp(-i\int_{t_{0}}^{t'}{\hat{H}(t'')dt''})$ and $\mathcal{T}$ is the time ordering operator.  For convenience we will choose one of the times in the two point function to be the initial time, i.e. $t_{0}=t$ and $t'\neq t$.  The two point function can be expressed in terms of the scalar field operator at time $t$ as
\begin{equation} \label{eq:twopointschrod}
<\!\!\Psi(t)|U^{\dagger}(t',t)\hat{\phi}(t,\vec{x}')U(t',t)\hat{\phi}(t, \vec{x})|\Psi(t)\!\!>
\end{equation}
where we have assumed above that $|0\!\!> = |\Psi(t)\!\!>$ is the state at the intial time $t$.  This equality is obtained by considering the solution to the Schr\"{o}dinger equation $|\Psi(t)\!\!> = \mathcal{T}U(t',t)|0\!\!>$ from which we get the desired result at $t'=t$.\\

After expanding to linear order in $\Delta t = t' - t$ in equation (\ref{eq:twopointschrod}) we obtain
\begin{equation}
<\!\!\Psi(t)|\hat{\phi}(\vec{x}')\hat{\phi}(\vec{x})|\Psi(t)\!\!> + i\Delta t<\!\!\Psi(t)|[\hat{H}(t),\hat{\phi}(\vec{x}')]\hat{\phi}(\vec{x})|\Psi(t)\!\!>
\end{equation}
Insert the resolutions of the identity in the basis of equation (\ref{eq:desitterbasis}) 
\begin{equation}
\hat{1} = \prod_{\vec{q} \epsilon \mathcal{H}}\int{da_{\vec{q}} |a_{\vec{q}}><a_{\vec{q}}|} 
\end{equation}
and perform the commutator to obtain $[\hat{H}(t),\hat{\phi}(\vec{x}')] = -i\hat{\pi}(\vec{x}')/a^2(t)$.  To make sure we have the right normalization we divide by the norm of the vacuum state.  This gives the following expression for the two point function up to linear order in $\Delta t$ in terms of the integrals over field configurations at some time:

\begin{equation} \label{eq:twopointfunc}
\frac{\int{\mathcal{D}\phi(x) \Psi^*(t)\phi(\vec{x}')\phi(\vec{x})\Psi(t) - \frac{i\Delta t}{a^2(t)}\Psi^*(t)\frac{\delta}{\delta \phi(\vec{x}')}(\phi(\vec{x})\Psi(t))}}{\int{\mathcal{D}\phi(x) \Psi^{*}(t)\Psi(t)}}
\end{equation}

To determine the exact wave functional $\Psi(t, \phi(x))$ that represents the conformal vacuum in the Schr\"{o}dinger picture we will compare the above expression for the approximate two point function to the explicit expression for the two point function in the conformal vacuum in de Sitter spacetime which can be found in the literature\cite{Birrell_Davies_1982}: 

\begin{equation}\label{eq:desittertwopoint}
\frac{1}{2}\int{\frac{d^3k}{(2\pi)^3k}\left(\frac{tt'}{C^2}\exp(-i k(t-t') +i\vec{k}\cdot(\vec{x}-\vec{x}'))\right)}
\end{equation}
where $t$ is the conformal time and $\vec{x}$ is the comoving coordinate.  Specifically, we will have to compare equation (\ref{eq:twopointfunc}) to (\ref{eq:desittertwopoint}) expanded to linear order in $\Delta t$.\\

To perform the integration over all fields in equation (\ref{eq:twopointfunc}) we go to the expansion (\ref{eq:desitterbasis}) of the field $\phi(x)$.  The measure becomes

\begin{equation}
\int{\mathcal{D}\phi(x)} \rightarrow \prod_{\vec{q} \epsilon \mathcal{H}}\int{da_{\vec{q}}}
\end{equation}
where $\mathcal{H}$ denotes the ``half-space'' of the momenta.  This is done so that we don't integrate over both $a_{\vec{q}}$ and $a_{-\vec{q}}$ as these variables are complex conjugates of each other for a real scalar field.  The integration is performed for the case $t=t'$, $\vec{x}\neq\vec{x}'$.  Integrating over all complex variables $a_{\vec{q}}$ for $\vec{q} \epsilon \mathcal{H}$ we obtain the expression for the two point function as
\begin{equation}
\int{\frac{d^3k}{(2\pi)^3} \left(\frac{1}{4f_{k}(t)} - \frac{\Delta t}{2a^2(t)} \left( i + \frac{g_{k}(t)}{f_{k}(t)} \right)\right) \exp(\vec{k}\cdot(\vec{x}-\vec{x}'))}
\end{equation}
Comparing the linearization of (\ref{eq:desittertwopoint}) we easily find the expression for $f_{k}(t)$ given in (\ref{eq:confvacuum}).  We can then use this $f_{k}(t)$ to solve for the $g_{k}(t)$ using equation (\ref{eq:fequation}).  The expressions obtained can then be verified to be solutions of (\ref{eq:gequation}).\\

Verifying that the two point functions are the same to zero'th and first order in $\Delta t$ is sufficient to establish the exact state functional because the two point function satisfies the homogeneous wave equation in a distributional sense (with cutoff)
\begin{equation}
g^{ab}\nabla_{a}\nabla_{b}<\Psi|\phi(x)\phi(x')|\Psi> = 0
\end{equation}
where $\nabla_{a}$ is the covariant derivative acting on the $x^c$ coordinate.  This can be checked by using the field equations for $\phi(x)$.  Therefore we have two solutions to the homogeneous wave equation in $x$ that agree at all $\vec{x}$ for $t=t'$ and whose first time derivatives agree at all $\vec{x}$ at $t=t'$.  By uniqueness of solutions to linear PDE's this means that the two point function calculated above is the two point function in the conformal vacuum.\\
  
The above procedure for identifying the conformal vacuum wave functional is easier than that found elsewhere \cite{Hill_1985} where it is done by renormalizing stress energy tensor to verify that the above wave functional is the conformal vacuum.\\

The response rate of an Unruh-Dewitt detector for a comoving observer in the conformal vacuum above has been calculated \cite{Birrell_Davies_1982} and is found to be equal to a thermal response in the comoving observer's proper time.
\begin{equation}
\dot{F}(\nu) = \frac{\nu}{2\pi(e^{2C\pi\nu}-1)}
\end{equation}
It is easily seen by substituting the solutions representing the conformal vacuum (\ref{eq:confvacuum}) in the expectation of the number operator (\ref{eq:particlenumber}) that the expectation of the number operator is not that of a thermal spectrum.  The expectation of the number operator and the response of an Unruh-Dewitt detector are therefore inconsistent.  

\section{A `natural' vacuum state in FRW}\label{sec:four}

The Hamiltonian for a Klein-Gordon field on a FRW metric with flat spatial sections (\ref{eq:flatfrw}) takes the form of a harmonic oscillator with time-dependent mass and frequency (\ref{eq:frwhamiltonian}).\\

Consider the real part of a single mode of the field.
\begin{equation}
H = \frac{1}{2}\left(\frac{p^2}{M(t)} + M(t)\Omega^2(t)x^2\right)
\end{equation}
where $x(t)$ is a real variable, $M(t) \equiv a(t)$ and $\Omega^2(t) \equiv \vec{k}^2 + \xi a^2(t)R(t)$, where $\xi$ is the coupling to the Ricci scalar of the field.  It will be demonstrated that this Hamiltonian can be put into the form of a Hamiltonian with \emph{fixed} mass and \emph{fixed} frequency by performing a canonical coordinate trasnformation on $x$ and $p$.\\

The new Hamiltonian must take the form
\begin{equation} \label{eq:newhamiltonian}
H' = \frac{1}{2m}P^2 + \frac{m\omega^2}{2}Q^2
\end{equation}
where $m$ and $\omega$ are constants and $Q$, $P$ are the new generalized coordinate and its conjugate momentum respectively.  To this end we write down the most general type 2 generating function that could possibly give rise to a new Hamiltonian of the form (\ref{eq:newhamiltonian}).  
\begin{equation}
G_{2} = f(t)xP + g(t)\frac{x^2}{2} + h(t)\frac{P^2}{2}
\end{equation}
The coordinate transformations are
\begin{eqnarray}\label{eq:canonicaltrans2}
p &=& \left(f(t) - \frac{h(t)g(t)}{f(t)}\right)P + \frac{g(t)}{f(t)}Q \\
x &=& \frac{Q - h(t)P}{f(t)}
\end{eqnarray}

Using the coordinate transformations and the equation $H' = H + \frac{\partial G_{2}}{\partial t}$ we find that the functions $f(t)$, $g(t)$ and $h(t)$ must satisfy
\begin{eqnarray}
\dot{h}(t) &=& \frac{1}{m} -\frac{f^2(t)}{M(t)} + m\omega^2h^2(t) \\
\label{eq:fequation2} \dot{f}(t) &=& \left(m\omega^2 h(t) - \frac{g(t)}{M(t)}\right)f(t) \\ 
\dot{g}(t) &=& m\omega^2f^2(t) - \frac{g^2(t)}{M(t)} - M(t)\Omega^2(t)
\end{eqnarray}
It is apparent that so long as $M(t)$ and $\Omega^2(t)$ are continuous and real then the above equations have real solutions.  In terms of the FRW spacetime, this means that the equations have solutions everywhere except at the big bang singularity.\\

The canonical transformations (\ref{eq:canonicaltrans2}) do not make sense at points where $f(t) = 0$.  However, we can integrate (\ref{eq:fequation2}) to obtain
\begin{equation}
f(t) = C\exp\left(\int{dt\left(m\omega^2h(t) - \frac{g(t)}{M(t)}\right)}\right)
\end{equation}
where $C$ is an integration constant.  Therefore, as long as the initial condition on $f(t)$ is chosen so that $C \neq 0$ then $f(t) \neq 0$ everywhere in the FRW universe, except at the big bang singularity where $M(t)$ is singular.\\

Thus we have a definition of particles, based on the energy levels of the harmonic oscillator, that is more natural in some sense than that used in other studies\cite{Krauss_2007}\;\cite{Greenwood_2010}\;\cite{Greenwood_2009}.  This again demontrates that the definition of particles based on the energy levels of the harmonic oscillator used elsewhere \cite{Krauss_2007}\;\cite{Greenwood_2010}\;\cite{Greenwood_2009} are physically meaningless and do not even represent the particle spectrum seen by some observer.

\section{Discussion}
It was demonstrated that the notion of particles, as defined by energy levels of the harmonic oscillator, correspond to our choice of coordinates on phase space.  By the principle of general covariance this definition is then, in general, unphysical and one should use a theoretical definition of particles that is independent of the coordinates chosen on phase space.  The Wightman function and consequently the response of an Unruh-Dewitt detector is such a definition that is coordinate independent.  The particle detector concept is thus more reliable than the energy levels of the harmonic oscillator concept and when these two definitions disagree one should trust the response of the Unruh-Dewitt detector.\\

Motivated by the above statements it was demonstrated that the conclusions found in other works \cite{Krauss_2007} \; \cite{Greenwood_2010} are unreliable due to the definition of particles used.  Using those authors' definition it has been demonstrated that one cannot identify a thermal or non-thermal state.  The response of the Unruh-Dewitt detector is the correct way to evaluate the response of an actual particle detector and this response has been found to disagree with the expectation of the number operator even in the standard case of de Sitter spacetime.\\

It was also demonstrated that there is a `more natural' definition of particles based on the energy levels of the harmonic oscillator than that used in the cited papers.  This was done by showing that one can put the Hamiltonian of a free Klein-Gordon field on FRW with flat spatial sections in the form of a sum of harmonic oscillators with time-independent frequency and mass.  This system has an obvious definition of particles akin to the usual definitions on flat spacetime that disagrees with both those found in the cited papers.  This further demonstrates the point that definitions of particles based on the energy levels of the harmonic oscillator have no physical meaning away from flat spacetime. \\

Further, we have demonstrated a new procedure for identifying the wave functionals in the functional Schr\"{o}dinger formalism of various states in the Heisenberg picture.  This procedure may be applied to find which states correspond to the ``ground state'' wave functional in various spacetimes where the scalar field takes the form of a harmonic oscillator with time dependent mass and frequency.  It is easier and more reliable than the method used by Hill \cite{Hill_1985} as that author compares the stress energy tensor operator calculated in the Heisenberg and Schr\"{o}dinger picture.  A renormalization must be done when calculating the stress energy tensor operator, so this makes the calculation less reliable.

\begin{acknowledgments}
We wish to acknowledge the support of an NSERC Discovery grant to C. C. Dyer.
\end{acknowledgments}
\appendix
\section*{Appendix: Relationship of particle number calculations to results in the literature \cite{Greenwood_2010}\;\cite{Greenwood_2009} \; \cite{Krauss_2007}}

Calculating the expectation of the number operator for some mode $a_{\vec{k}}$ as done above can be easily related to the procedure found in other works \cite{Greenwood_2010}\; \cite{Greenwood_2009}\;\cite{Krauss_2007}.  In these works the expectation of the number operator is found by using directly the functions representing the n-particle states of some mode.  These functions are denoted $H_{n}(t, a_{\vec{k}})$ and are the Hermite polynomials multiplied by a Gaussian factor like $\exp(-h(t)a_{\vec{k}}a_{-\vec{k}})$ where $h(t)$ denotes a complex function of time.  We assume that these functions are orthonormal
\begin{equation} \label{eq:orthonormality}
\int{da_{\vec{k}}H^{*}_{m}(t, a_{\vec{k}})H_{n}(t, a_{\vec{k}})} = \delta_{nm}
\end{equation}
They are also eigenstates of the number operator
\begin{equation} \label{eq:eigenstate}
N_{\vec{k}}H_{n}(t, a_{\vec{k}}) = nH_{n}(t, a_{\vec{k}})
\end{equation}
They form a complete basis of functions so we can write the wave function for a single mode ($a_{\vec{k}}$) at some time in terms of these particle basis state at that time
\begin{equation}\label{eq:completeness}
\Psi(t, a_{\vec{k}}) = \sum_{n}{c_{n}(t)H_{n}(t, a_{\vec{k}})}
\end{equation}
It can be easily shown using the orthonormality (\ref{eq:orthonormality}), the eigenstate property (\ref{eq:eigenstate}) and the completeness (\ref{eq:completeness}) that the following equality holds.
\begin{equation}
<N_{\vec{k}}> = \int{da_{\vec{k}}\Psi^*(t,a_{\vec{k}})N_{\vec{k}}\Psi(t, a_{\vec{k}})} = \sum_{n}{n|c_{n}(t)|^2}
\end{equation}
Therefore, the calculation of the expectation of the number operator above is equivalent to that found in the referenced literature \cite{Greenwood_2010} \; \cite{Greenwood_2009} \; \cite{Krauss_2007} for a single mode.  It is important to note that the multiparticle states are not necessarily solutions of the Schr\"{o}dinger equation which is why the coefficients in equation (\ref{eq:completeness}) are functions of time.

\bibliographystyle{plain}
\bibliography{myrefs}
\nocite{*}

\end{document}